\documentclass{appolb}
\usepackage[pdftex]{graphicx}
\usepackage{cite,color}
\def\Im{{\rm Im}}

\begin{document}
\title{Continuous decoupling and freeze-out%
  \thanks{Presented at  the IV International  Workshop on Correlations
    and Femtoscopy, Krakow, Poland - September 2008} %
}
\author{J\"orn Knoll
\address{GSI, Helmholtzzentrum f\"ur Schwerionenforschung, 64291
  Darmstadt, Germany}
}
\maketitle
\begin{abstract}
  The  decoupling and  freeze-out of  energetic nuclear  collisions is
  analysed in terms of transparent semi-classical decoupling formulae.
  They provide a smooth  transition and generalise frequently employed
  instantaneous freeze-out  procedures.  Simple relations  between the
  damping  width  and  the  duration  of the  decoupling  process  are
  presented and the implications on various physical phenomena arising
  from  the expansion  and  decay dynamics  of  the highly  compressed
  hadronic  matter generated  in  high energy  nuclear collisions  are
  discussed.

\end{abstract}
  
\section{Introduction}

Dynamically expanding systems may pass various stages, where different
dynamical concepts or description levels are appropriate.  For nuclear
collisions a  possibly formed quark  gluon plasma (QGP) converts  to a
dense  hadronic medium.   Subsequently the  chemical  components (i.e.
the  abundances of  the different  hadrons) decouple  and  finally the
system kinetically freezes out  releasing the particles that reach the
detectors.

All such transitions share  that they need quite some time, be  it in order to
cope  with strong rearrangements  of the  matter, in  part accompanied  with a
significant change in entropy density and corresponding release of latent heat
(e.g.   for  the QGP  $\to$  hadron matter  transition),  or  simply that  the
processes happen probabilistically.  Some  of these transitions can adequately
be formulated  in terms  of micro- or  macroscopic transport  equations, other
ones such  as e.g.  phase transitions or  composite-particle formation involve
changes of degrees of freedom,  which are complicated in microscopic terms and
mostly not yet well formulated.   Therefore in many cases recipes are applied,
mostly  of instantaneous  nature.  These  concern coalescence  pictures, which
combine  nucleons  to composite  nuclei  at  freeze-out \cite{Sato:1981ez}  or
coalesce  quarks  to   hadrons  in  the  deconfinement-confinement  transition
\cite{Biro:1998dm}.  Also the general freeze-out is mostly treated as a sudden
transition  happening  at  a  suitably  chosen  three  dimensional  freeze-out
hyper-surface in space-time \cite{Cooper74}.  In many cases such prescriptions
violate general  principles as detailed balance,  unitarity, conservation laws
or   entropy   requirements  \cite{flavor_kin:Barz88,flavor_kin:Barz90}.    In
particular  the instantaneous  freeze-out picture  is widely  used  to analyse
nuclear  collision data  in terms  of thermal  models.  The  achieved  fits in
temperature, chemical potentials and parametrised flow effects to the observed
particle abundances and kinetic spectra,  are then frequently used as measured
data that are taken as clues on the physics of the collision dynamics, cf.  A.
Andronic,   and   W.   Broniowski   on   this   workshop   or  Refs.    \cite{
Cleymans:2005xv, Andronic:2005yp}. 

In this contribution I'll  present some simple analytical considerations which
illustrate the  space-time dynamics of transition processes  in didactic terms
at the  example of the  freeze-out. The results  are well in line  with recent
progress reported  by Y.   Sinyukov and  S.  Pratt at  this meeting,  see also
\cite{Sinyukov:2002if,  Akkelin:2008eh, Pratt:2008sz,  Pratt:2008qv}.  Earlier
attempts  towards  a  continuous  freeze-out  description as  derived  by  the
Brazilian group  \cite{Grassi:1994nf, Grassi:1994ng} and  later efforts \cite{
Csernai:1997xb,  Magas:1999yb, Molnar:2005gx}  directly focussed  on  a global
decoupling  scheme for  fluid-dynamic approaches.   Here we  investigate these
processes   micro-dynamically   similar  to   the   conceptual  progress   and
clarifications given in \cite{Sinyukov:2002if, Akkelin:2008eh} on the basis of
classical kinetics.

\section{Decoupling formulae}
The decoupling  properties of  an observed particle  are given  by its
interaction with the particles in the source. The latter is encoded in
the corresponding current-current correlation function or polarisation
function  $\Pi$. It provides  the following  local decoupling  rate in
four phase space\cite{Knoll:2008sc,Knoll:2008}
\begin{eqnarray}
\label{eq:emissivity-local}
 \frac{dN(x,p)}{ d^4pdtd^3x}
&=&\frac{1}{(2\pi)^4}
\underbrace{\Pi^{\rm gain}(x,p) A(x,p)}_{C^{\rm gain}(x,p)}
\ {\cal P}_{\rm escape}(x,p).
\end{eqnarray}
in    the   context   of    the   gradient    expanded   Kadanoff-Baym
equations\cite{KB1962,Ivanov2000}.  Here  the first two  factors (gain
rate encoded  in $\Pi^{\rm gain}$  times local spectral  function $A$)
determine  the   gain  part  of  the  local   collision  term  $C^{\rm
  gain}(x,p)$\cite{Ivanov2000}.    The  last  factor   ${\cal  P}_{\rm
  escape}(x,p)$  captures the probability  that particles,  created or
scattered at space-time  point $x$ into a momentum  $p$, can escape to
infinity  without further  being  absorbed  by the  loss  part of  the
collision term.  This formulation  thus restricts the emission zone to
the  layer of  the last  interaction.  Semi-classically  in  the small
damping width limit one obtains
\begin{eqnarray}
\label{eq:Pescape}
{\cal P}_{\rm escape}(x,p)&=&{\rm e}^{-\chi(x,p)},\quad \mbox{where}\quad
\chi(x,p)=\int_{(x,\vec{p})}^{\infty}\Gamma(x',p')dt'
\end{eqnarray}
with    $\Gamma(x',p')=-\Im\Pi^{\rm    R}(x',p')/p'_0$.    The    time
integration defining the optical depth $\chi$ runs along the classical
escape path  starting at $(x,\vec{p})$,  the latter determined  by the
real  part  of  the   retarded  polarisation  function  $\Pi^{\rm  R}$
\cite{Knoll:2008sc,Knoll:2008}.   The  above  rate  causes  drains  in
particle number and energy and a recoil momentum from the source which
in a fluid dynamical description lead to
\begin{eqnarray}\label{fluid-drain}
\partial_{\mu}
{j_{\alpha,\rm fluid}^{\mu}(x)\choose T^{\mu\nu}_{\rm fluid}(x)}&=&
-\sum_a\int\! d^4 p\ 
{e_{a\alpha}\!\choose p^{\nu}}\frac{d\,N_a(x,p)}{d^4xd^4p},
\end{eqnarray}  
resulting  from  the  dissipative  part of  the  underlying  transport
equations  (here $a$  labels the  different particles  and  $\alpha$ a
conserved current).   For small  spectral width all  spectral strength
$A(x,p)$ will  be guided towards  the detector with  on-shell momentum
$\vec{p}_A$ providing the following detector yield
\begin{eqnarray}
\label{eq:emissivity-detector}
\hspace*{-1em} \frac{dN_a(p_A)}{ d^3p_A}
&=&\int\! \frac{d^4 x  d^4p}{(2\pi)^4}\  
\Pi_a^{\rm gain} 
 A_a\ {\cal P}_{\rm escape}
\left(\!\frac{\partial\vec{p}_A}{\partial\vec{p}}\!\right)^{\!\! -1}
\!\delta^3(\vec{p}-\vec{p}(x,\vec{p}_A)).
\end{eqnarray}
Here  $\vec{p}_A(x,\vec{p})$  and  $\vec{p}(x,\vec{p}_A)$  denote  the
corresponding mapping of the  local momentum $\vec{p}$ to the detector
momentum  and  its inverse,  respectively.   The corresponding  Jacobi
determinant accounts for the  focussing or defocusing of the classical
paths  due to  deflections.

In  thermal  equilibrium  the  source function becomes
\begin{eqnarray}\label{PiThermal}
\Pi_a^{\rm gain}(x,p)=-2\ f_{\rm th}(x,p^0)\ \Im\Pi_a^{\rm R}(x,p)
=f_{\rm th}(x,p^0)\ 2 p^0\ \Gamma_a(x,p), 
\end{eqnarray}
where $\Gamma_a(x,p)$ is the local damping width of particle $a$. This
property  leads to  quite  some compensation  effect,  which is  frame
independent. Namely, for large source extensions the integral over the
$\Gamma$-dependent damping  factors in (\ref{eq:emissivity-detector}),
which define  the visibility  probability $P_t$,  equates to
unity
\begin{eqnarray}\label{GammaUnity}
\int_{-\infty}^{\infty} dt\ \underbrace{\Gamma(t)\ \textstyle
{\rm e}^{-\chi(t)}}_
{=\ P_t(t)}
&=&1,\quad\mbox{where}\quad
\chi(t)=\int_t^{\infty} dt' \Gamma(t'),
\end{eqnarray}
if integrated along any path leading from the {\em opaque} interior to
the  outside.   This compensation  is  independent  on the  structural
details  of  $\Gamma$  and  on  the classical  paths  leading  to  the
detector, along which the decoupled particles are accumulated.
When rates  drop smoothly in time the  visibility probability $P_t(t)$
achieves its maximum at
\begin{eqnarray}\label{emission-max-t}
\left[\frac d{dt}{\Gamma}(t)+ \Gamma^2(t)\right]_{t_{\rm max}}&=&0,
\quad\quad\quad\mbox{where}\quad
P_t(t_{\rm max})\approx\Gamma(t_{\rm max})/
{\rm e}.
\end{eqnarray}
The   corresponding   decoupling   duration   $\Delta   t_{\rm   dec}$
approximately  follows  from   the  normalisation  of  the  visibility
function $P_t$  through a kind  of decoupling uncertainty  relation ${
  \Gamma(t_{\rm max})}\ \Delta t_{\rm dec}\approx {\rm e} $.

In the limit that $P_t(t)$  can be replaced by a $\delta$-function one
recovers  an improved  Cooper-Frye\cite{Cooper74} formulae,  where the
freeze-out   hypersurface   is  no   longer   globally  defined,   but
individually  by the  posed  detector momentum  \cite{Sinyukov:2002if}
through   the   features    of   $\Gamma(x,p)$   at   peak   condition
(\ref{emission-max-t}).
\begin{table}[b]
\centerline{\fbox{
\begin{tabular}{rcc}
{\bf 100 GeV Au + Au} &decoupl. time&vol. growth\\
phase transition \cite{flavor_kin:Barz88,flavor_kin:Barz90}:   
& 6 - 10  fm/c  &$>    5$\\
chemical freeze-out:&$>   5    $ fm/c&$>    4$\\
kinetic freeze-out: &$>   8    $ fm/c&$>    6$\\
{\bf CMB early universe}\cite{Mukhanov2005}:& 
$Z=[1300-800]$& $(13/8)^3=4.3$
\end{tabular}
}}
\caption{Typical decoupling durations and volume growths.}\label{Tab1}
\end{table}
\section{Analytic model considerations}

In  Ref.\ \cite{Knoll:2008sc} various  consequences of  the continuous
decoupling formalism  are discussed.   As an illustration  we consider
the competition  between chemical and kinetic  (thermal) freeze-out of
slow particles escaping from a spherically expanding uniform fireball.
The emission  is than essentially from a  time-like hypersurface. Both
processes go  with a  different pace as  a function of  density and/or
temperature during the expansion,  since inelastic processes drop much
faster than  the elastic scattering processes,  the latter essentially
determining the kinetic rates.

For example a  fireball evolution with a freeze-out  radius of $R_{\rm
  dec}\approx 6$ fm and  collective velocity $\dot{R}_{\rm dec}= 0.5 $
fm/c  leads to a  decoupling peak  at $t_{\rm  max}=12$ fm/c  for both
types of  freeze-out.  The  damping widths at  decoupling peak  are as
large as $\Gamma^{\rm chem}_{\rm max}=\Gamma^{\rm kin}_{\rm max}= 0.5$
c/fm $\approx 100$ MeV providing the values given in Table \ref{Tab1}.
The  typical duration  of  a  decoupling process  is  thereby of  less
importance.

Rather the robust  feature is the relative volume  growth during which
the  system  decouples,  which  is  mostly beyond  half  an  order  of
magnitude. During this time the thermodynamic properties of the system
can  significantly  change thereby  influencing  the  spectrum of  the
observed particles. Even  more robust is the behaviour  of the overall
damping  rate $\Gamma$  of  the decoupling  particle.   Coming from  a
completely opaque  zone where the  damping is strong, it  decreases by
about a factor ${\rm e}^ {\rm e}\approx 15$ with values between begin,
peak and end of  the freeze-out of notably $\Gamma_{\rm i}:\Gamma_{\rm
  max}:\Gamma_{\rm f}=390:100:26$~MeV for  the above nuclear collision
scenario.

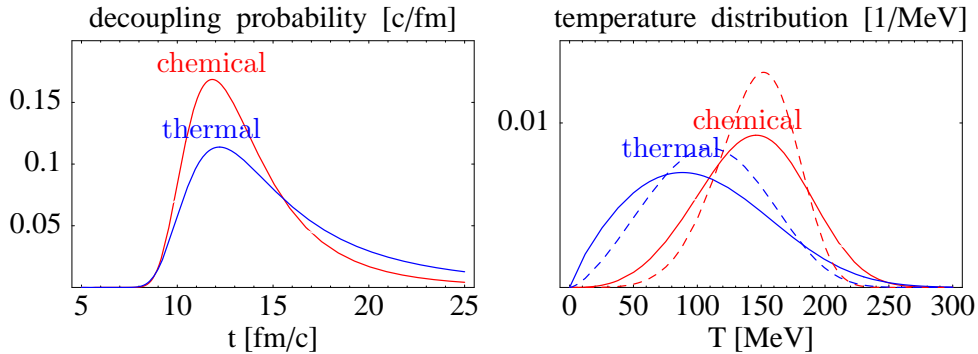
\begin{figure}[t]
\unitlength0.01\textwidth
\begin{picture}(100,37)
\put(-1,0){\includegraphics[width=0.6\textwidth]{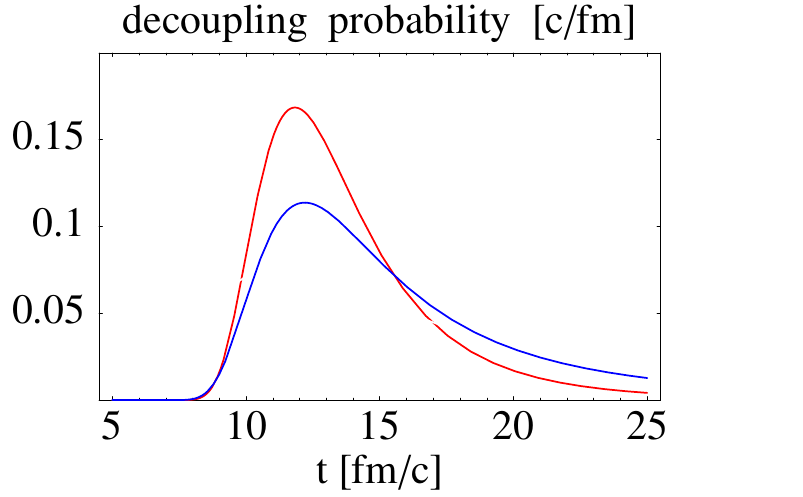}}
\put(21,31){\makebox(0,0){\color{red}chemical}}
\put(21,24){\makebox(0,0){\color{blue}thermal}}
\put(50,0){
\put(0,0){\includegraphics[width=0.6\textwidth]{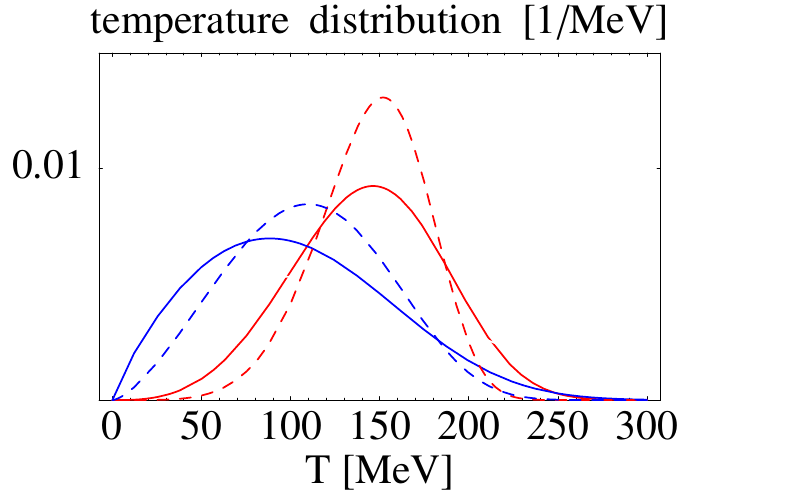}}
\put(27,25){\makebox(0,0){\color{red}chemical}}
\put(19,22){\makebox(0,0){\color{blue}thermal}}}
\end{picture}
\caption{Decoupling probability  $P_t(t)$ as a function  of time (left
  panel)  and the resulting  temperature distributuions  (right panel)
  the  latter for  two simple  EoS with  adiabatic  index $\kappa=1.5$
  (full  lines)  and $\kappa=4/3$  (dashed  lines)  for the  schematic
  chemical  and thermal  freeze-out  scenarios discussed  in the  text
  \cite{Knoll:2008sc}.}
  \label{emission-brilliance}
\end{figure}
 
Depending on  the underlying EoS  the thermodynamic properties  of the
matter can  therefore significantly change during  the decoupling time
window.      The     resulting     distributions    in     temperature
$P_T(T)=P_t(t)(dT(t)/dt)^{-1}$  as shown  in the  right panel  of Fig.
\ref{emission-brilliance} for two normally  behaved example EoS show a
significant spread  in the  $T$ distributions.  Remarkable  is further
that  although both  time distributions  $P_t$ peak  at the  same time
(left panel  of Fig.  \ref{emission-brilliance}),  the slower decrease
in kinetic  rates leads  to a considerably  downward shifted  and much
broader $T$  distribution for thermal  freeze-out compared to  that in
the chemical case.

\section{Summary and perspectives}

For any  observed probe  the structural properties  of the  source are
encoded in the current-current  correlation function defining the gain
part of  the polarisation function  $\Pi$ (or self-energy).   It knows
about resonances  and other  structural properties that  influence the
emitted particle.  Only penetrating probes, which suffer no distortion
and absorption in the matter  (${\cal P}=1$), have an undisturbed view
on  the   source.   Strongly   interacting  probes,  though,   have  a
significantly  reduced  view.  In  the  opaque  limit most  structural
effects are  wiped out and one  is rather left  to observe statistical
features,  i.e.  temperature  effects, of  the source.   In particular
kinematical fingerprints from decays of resonances with lifetimes less
than  the decoupling  duration will  become invisible  in  the kinetic
spectra  of the  probe, deferring  to  include such  decay modes  into
statistical model descriptions.

The  here  discussed  decoupling  features  are  generic  and  apply  to  many
dynamically  expanding  systems.    This  includes  the  microwave  background
radiation released  during the early universe evolution.   For applications to
nuclear collisions the here diagnosed long decoupling and freeze-out times are
both a  challenge but  also a chance:  a chance  to map out  the thermodynamic
properties of  the expanding collision  zone during the freeze-out  of various
probes.  An  observation of  a quite narrow  distribution in  temperature, for
example,  could  point  towards  effects  that  significantly  slow  down  the
temperature  drop during  expansion, and  this way  provide hints  towards the
underlying equation of state  or possible phase transition effects.  

In summary I see  a need to reanalyse nuclear collisions data  in the light of
the results  and discussions  given here.  Promising  steps towards  this goal
were   presented   at  this   meeting   by   Y.    Sinyukov  and   S.    Pratt
\cite{Akkelin:2008eh, Pratt:2008sz}.   In those hybrid  model calculations the
entire  decoupling stage  is treated  within kinetic  transport.   These model
calculations  do  not  only  confirmed  the  here  advocated  long  freeze-out
durations  (above  10  fm/c).  At  the  same  time,  contrary to  earlier  HBT
interpretations   that   inferred   extremely   short   decoupling   durations
\cite{Adamova:2002wi}, these long durations  emerged well conform with $R_{\rm
out}/R_{\rm side}$ close to unity  \cite{Pratt:2008qv} as observed in the data
at the CERN SPS and at RHIC,  this way providing a solution to the alleged HBT
puzzle discussed on this workshop,  cf.  also the corresponding HBT reviews of
\cite{Csorgo:2005gd, Padula:2004ba}.
\def\baselinestretch{0.85}\normalsize         
\bibliography{References}
\bibliographystyle{elsart-num}

\end{document}